\documentstyle[pra,aps,twocolumn,epsfig]{revtex}

\begin{document}

\title{Dynamical Cooling of Trapped Gases I: One Atom Problem}

\author{Luis Santos$^{\dag}$ and Maciej Lewenstein$^{\ddag}$}

\address{\dag Departamento de F\'\i sica Aplicada, Universidad de
Salamanca, Plaza de la Merced s/n, 37008 Salamanca, Spain \\
\ddag Institut f\"ur Theoretische Physik, Universit\"at Hannover,
 Appelstr. 2, D-30167 Hannover,
Germany}

\maketitle

\begin{abstract}
We study the laser cooling of one atom in an harmonic trap beyond the Lamb-Dicke regime. By using
sequences of laser pulses of different detunings we show that the atom can be confined into just
one state of the trap, either the ground state or an excited state of the harmonic potential. The
last can be achieved because under certain conditions an excited state becomes a dark state. We
study the problem in one and two dimensions. For the latter case a new cooling mechanism is
possible, based on the destructive interference between the effects of laser fields in different
directions, which allows the creation of variety of dark states. For both, one and two dimensional
cases, Monte Carlo simulations of the cooling dynamics are presented.
\end{abstract}

\section{introduction}
\label{sec:Intro}

Atomic laser cooling has become recently one of the most active research fields in atomic physics
\cite{Aspect}. Several techniques \cite{VSCPT,Raman} have been developed to cool free atoms up to
fractions of the recoil temperature (i.e. the temperature associated with the recoil of a single
photon). The sub-recoil laser-cooling methods are based on the crucial concept of dark states.
These states cannot absorb the laser light, and therefore cannot be excited, but however can
receive population via incoherent pumping, i.e. via spontaneous emission, and therefore these
states behave as trapping states. The dark-state is a general concept, and can be obtained by
several mechanisms, for example the destructive interference between the absorptions from the two
ground states of a three level $\Lambda $-system, as in Velocity Selective Coherent
Population Trapping (VSCPT) \cite{VSCPT}. On the other hand, laser cooling of trapped atoms to the
ground state of a trap has been achieved for single trapped ions using sideband cooling
\cite{SDBC} in the Lamb-Dicke limit (LDL) \cite{Stenholm}. This limit corresponds to the case
in which the size of the ground state of the trap $a_{0}$ is very small compared with the
wavelength of the applied laser $\lambda$. More rigorously, LDL occurs  when $\eta \ll 1$,
where $\eta$ is the so-called Lamb-Dicke parameter  defined as $\eta = 2\pi a_{0}/\lambda$. Sideband
Cooling method only works properly in this limit, and also requires the levels of the trap to be
well resolved, which is achieved in the so-called strong-confinement limit, i.e. when
$\gamma <\omega$, where $\omega$ is the trap frequency and $\gamma$ is the spontaneous emission
rate (the level width). This last condition can be easily fulfilled by using a metastable excited
state and quenching its population using a third fast-decaying level \cite{Marzoli}. However, the
problem appears in the regime beyond the LDL, in which the recoil of the spontaneously emitted
photons provides a heating mechanism, and therefore new cooling strategies must be proposed. The
most effective techniques have been those originally designed to cool free atoms up to subrecoil
temperatures, i.e. dark-state cooling techniques \cite{Dum}. A recent paper \cite{Morigi}, proposes
a new cooling scheme for trapped atoms beyond the LDL which combines the ideas of sideband and
Raman \cite{Raman} cooling and dark-state cooling, to confine a single trapped atom in the ground
state of the trap. 

In this series of  papers, we study laser cooling techniques similar to that of \cite{Morigi} which
allow the confining of an atomic sample into just one state of the trapping potential, either the
ground state or an excited state. This first paper is devoted to the one-atom problem while the
second one \cite{Uns} analyses the new physics appearing for the many-atom case. The one-atom
problem is interesting  as such in the context of quantum computation with ions \cite{Cirac}, and
also in the production of non-classical states of motion of single atoms or ions \cite{Cirac2}.
Also the one-atom problem is a good starting point to  understand  clearly  the cooling processes
in the many-atom case, which we discuss in detail in \cite{Uns}; therefore this paper is a good
first step to analyse the possibilities to achieve the Bose-Einstein Condensation \cite{BEC} with
all-optical means. 

The paper is organized as follows. In Sec.\ \ref{sec:Model} the cooling scheme is described, as
well as the difficulties to cool beyond the LDL. In Sec.\ \ref{sec:1D} we study the
one-dimensional case, analysing the cooling into the ground state of the trap (basically
summarizing \cite{Morigi}) in Sec.\ \ref{sec:1DG}. The one-dimensional cooling into an
excited of the trap is studied in Sec.\ \ref{sec:1DE}, where the dark-state conditions are
presented. In section Sec.\ \ref{sec:2Dnew} and Sec.\ \ref{sec:2DRE} the new physics which appears
in the two-dimensional case is considered. In Sec.\ \ref{sec:2Dint} a new interference
mechanism between the lasers in different directions is presented. In Sec.\ \ref{sec:2DG}
the  cooling into the ground state is analysed, whereas in Sec.\ \ref{sec:2DE}, we employ
the dark-state conditions of Sec.\ \ref{sec:1DE} and also the interferential dark-states of
Sec.\ \ref{sec:2Dint}, to confine the atom in an excited two-dimensional state. We conclude in
Sec.\ \ref{sec:Conclusions}. 

\section{Model}
\label{sec:Model}

\begin{figure}[ht]
\begin{center}\
\epsfxsize=5.5cm
\hspace{0mm}
\psfig{file=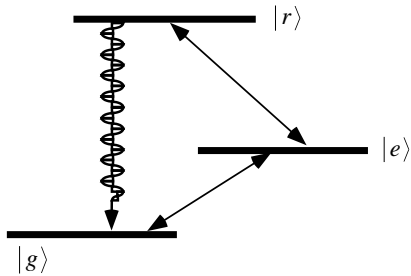,width=5.5cm}\\[0.3cm]
\caption{Internal atomic configuration considered throughout the paper.}
\label{fig:1}
\end{center}
\end{figure}

We consider laser cooling in a three-level $\Lambda$-system as that presented in Fig.\ \ref{fig:1}.
The level $|g\rangle$ is the ground state, $|e\rangle$ is a metastable state, and $|r\rangle$ is a
third fast-decaying auxiliary state. A laser, or stricktly speaking two lasers excite coherently the
resonant Raman transition $|g\rangle\rightarrow|e\rangle$ (with some associated effective Rabi
frequency $\Omega$), while a repumping laser in or off-resonance with the transition
$|g\rangle\rightarrow|r\rangle$ pumps optically the atom into $|g\rangle$. With this three level
scheme, one obtains an effective two-level system with an effective spontaneous emission rate
$\gamma$, which can be controlled by varying the intensity or the detuning of the repumping laser
\cite{Marzoli}. The atom is assumed to be confined in an harmonic (isotropic) potential of
frequency $\omega$. We denote the states of the harmonic oscilator as $|n\rangle$, or $|n_{x},
n_{y}\rangle$ when treating the two-dimensional case. We assume $\gamma <\omega$, i.e. we suppose
the so-called strong-sonfinement limit, which is also called Festina-Lente limit, and turns to be a
crucial point in the many atom case  in order to avoid the reabsorption problem
\cite{Uns,Reabsorption}. We also assume $\Omega < \gamma $ and $\Omega ^{2}/\gamma \ll \omega$,
which allows  to adiabatically eliminate the excited state $|e \rangle $ in the expressions below.

In the Lamb-Dicke limit (LDL), i.e. when the dimensions of the trap are very small compared with
the laser wavelength $\lambda$, it is possible to use standard Sideband Cooling techniques to pump
the atom into the ground state of the harmonic trap. In this cooling mechanism the laser is assumed
red-detuned, with detuning $\delta =-\omega$. Then, the absorption of a laser photon induces the
transition $|g,n\rangle \rightarrow |e,n-1\rangle$. In the LDL, $\eta^{2} =E_{R}/\hbar\omega \ll 1$
, where $E_{R}$ is the recoil energy. Then, in the spontaneous emission the photon recoil is not
large enough to produce a jump between the states of the harmonic trap; therefore only the
transition $|e,n-1\rangle \rightarrow |g,n-1\rangle$ is possible. Hence, the global effect of a
complete absorption-spontaneous emission cycle is a transition $|g,n \rangle \rightarrow
|g,n-1\rangle$. This process is repeated until the particle reaches the ground state of the trap,
where it can no longer be excited by the laser, i.e. $|g,0\rangle$ is a dark state. The latter
statement is only true if we operate in the regime $\gamma <\omega$.

Beyond the LDL, $E_{R}\ge \hbar \omega$, and hence in the spontaneous process, the atom undergoes
transitions $|e,n-1\rangle \rightarrow |g,n-1\pm n'\rangle$ where $n'$ ranges from $0$ to ${\cal O}
(\eta ^{2})$. This process introduces a heating of the atomic distribution which prevents the
confinement of the atom into the ground state of the trap. In this paper we study different
techniques which allow us to overcome this problem, recovering the results of usual Sideband
Cooling, and even obtaining new physical effects.

\section{One Dimension}
\label{sec:1D}

\subsection{Cooling into the Ground State of the Trap}
\label{sec:1DG}

\subsubsection {Cooling Mechanism}
\label{sec:1DGCM}

The previous paper \cite{Morigi} studied the problem of laser cooling of a single trapped atom
beyond the LDL into the ground state of a one-dimensional trap. In this section we briefly review
that paper, and discuss a cooling mechanism which avoids the heating effects of the usual Sideband
Cooling by employing the two step process:

\begin{itemize}
\item  In the first step the atoms are confined in the levels $n \le \eta^{2}$. This is achieved by
using pulses with detunings $\delta =-\hat \eta ^{2} \omega $, where $\hat\eta ^{2}$ is the closest
integer to  $\eta ^{2}$. The absorption of a laser photon produces a transition $|g,n\rangle
\rightarrow |e,n-{\cal O}(\eta ^{2})\rangle$, and the spontaneous emission $|e,n-{\cal O}(\eta
^{2})\rangle \rightarrow |g,n-{\cal O}(\eta ^{2})\pm n'\rangle$, where $n'$ ranges from $0$ to
${\cal O} (\eta ^{2})$. As a result, in average some energy is lost in each cycle. Obviously, the
states with $n<\eta ^{2}$ remain dark, and therefore the atoms are finally confined in these
states. Due to this, we call these pulses confinement pulses.

\item Using pulses of selected frequencies one can empty the trap levels $(|g,1 \rangle , |g,2
\rangle , \dots)$ without affecting the population of $|g,0 \rangle $. Then, in each cooling cycle
there is some probability to pump into $|g,0 \rangle $. There is also some probability of heating,
and therefore we must repeat the confinement step, and so on. We call the pulses with selected
frequency fine-cooling pulses.

\end{itemize}

As the  result of the cooling  process, the atom can be confined in the ground state of the trap,
as in the usual Sideband Cooling. Since the present cooling mechanism is based on sequences of
different pulses, we call this cooling method Dynamical Cooling.

\subsubsection {Rate Equations}
\label{sec:1DGRE}

The dynamics of the system can be well described, after adiabatic elimination of the $|e \rangle $
states, by the following rate equation:

\begin{equation}
\dot N_{m}=\sum_{n=0}^{\infty}\Gamma_{m \leftarrow
n}N_{n}-\sum_{n=0}^{\infty}\Gamma_{n
\leftarrow m}N_{m},
\label{rateeq1d}
\end{equation}

where $N_{n}$ is the probability of the atom to be in the level $|n \rangle$ of the trap. The
transition rates are:

\begin{eqnarray}
&& \Gamma_{n \leftarrow m}=\frac {\Omega ^{2}}{2\gamma}\int_{0}^{2 \pi}d\phi
\int_{0}^{\pi} \sin{\theta} d\theta {\cal W}(\phi , \theta )\nonumber \\
&& \times \left |\sum_{l=0}^{\infty}
\frac {\gamma \langle n|e^{ik \sin{\theta} \cos{\phi} x}
|l\rangle \langle l|e^{ikx}|m\rangle}{[\delta-\omega (l-m)]+i\gamma}\right
|^{2},
\label{rates1d}
\end{eqnarray}

where $k$ is the laser wavevector and ${\cal W}$ is the angular dipole pattern of the spontaneously
emitted photons. Using closure relations, assuming a laser detuning $\delta =s \omega $ with $s$ an
integer value, and neglecting the non-resonant terms, it is possible to write the rates to empty
the level $|m\rangle $ (called hereafter empty rates) in a very simple form in terms of the
Franck-Condon factors:

\begin{eqnarray}
&& \Gamma_{m}=\sum_{n\not= m}\Gamma_{n \leftarrow m}\nonumber \\
&& \qquad \simeq \frac{\Omega^{2}}{2\gamma}
\cases {|\langle m+s|e^{ikx}|m\rangle|^{2}, &if $m+s>0$;
 \cr 0, &otherwise. \cr}
\label{erates1d}
\end{eqnarray}

\subsubsection{Numerical results}
\label{sec:1DGNR}

\begin{figure}[ht]
\begin{center}\
\epsfxsize=7.5cm
\hspace{0mm}
\psfig{file=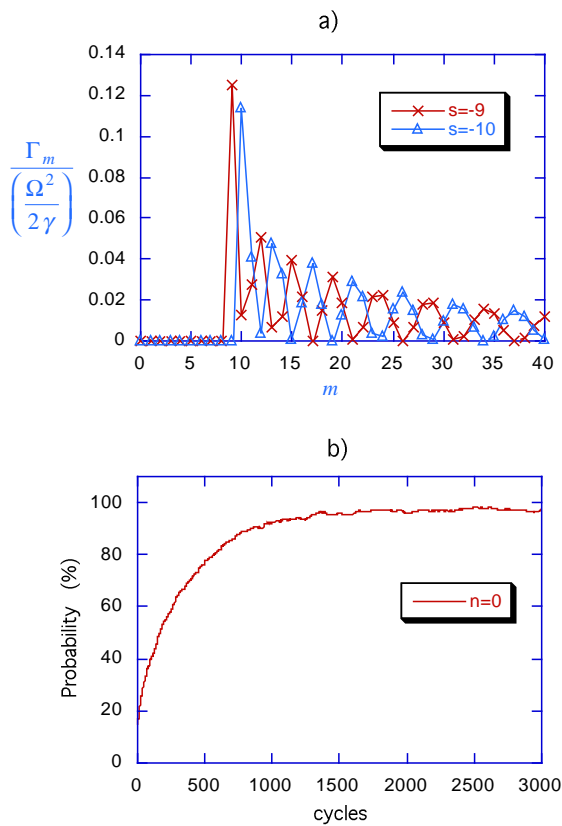,width=7.5cm}\\[0.3cm]
\caption{(a) Empty rates as a function of the trap level, for $\eta =3.0$, 
$\gamma=0.01\omega$, and for the case of a pulse with $\delta=s\omega$, with $s=-9, -10$. (b)
Dynamics of the occupation probability of the level $n=0$ as a function of the applied cycles,
starting from the thermal state with the mean number $6$. Each cycle consists on a sequence of four
pulses of detunings $\delta_{1,2,3,4}=-9\omega, 0, -10\omega, -\omega $ and durations
$T_{1,2,3,4}=2\gamma/\Omega ^{2}$.}
\label{fig:2}
\end{center}
\end{figure}

Monte Carlo simulations of Eq.\ (\ref{rateeq1d}) yield similar results to those of \cite{Morigi}.
In Fig.\ \ref{fig:2}b the cooling dynamics for the case of $\eta =3.0$, $\gamma=0.01 \omega$ is
shown, where the initial distribution was taken to be thermal with the  mean $\langle n \rangle
=6$. We use a sequence of four pulses of detunings $\delta_{1,2,3,4}=-9\omega, 0, -10\omega,
-\omega $ and durations $T_{1,2,3,4}=2\gamma/\Omega ^{2}$ to confine the atom with almost total
probability in the ground state of the trap. Pulses $1$ and $3$ are confining pulses, and $2$ and
$4$ are fine-cooling pulses. The reason of applying two different confinement pulses slightly
detuned one respect to the other, is the oscilatory character of the Franck-Condon factors, which
produces, as observed in Fig.\ \ref{fig:2}a, that the empty rates have quasi-zero minima for each
detuning. By applying two sligthly detuned lasers this problem is avoided because the states that
one laser does not empty the other laser does, and vice versa. The same reasoning explains the use
of two fine-cooling pulses.

\subsection{Cooling into an Excited State of the Trap}
\label{sec:1DE}

\subsubsection{Condition to obtain a Dark Excited State of the Trap}
\label{sec:1DEC}

In the present section we present a new mechanism that allows the confining of the atom into an
excited state of the trap. This cooling mechanism is based on the properties of the Franck-Condon
factors appearing in Eq.\ (\ref{erates1d}), which are of the form:

\begin{eqnarray}
&& \langle m+s|e^{ikx}|m\rangle = i^{2m+s} e^{-\eta ^{2}/2}\sqrt{m!(m+s)!},
\nonumber \\
&& \times
\sum_{l=0}^{\min{[m,m+s]}}\frac{(-1)^{l}\eta^{2m+s-2l}}{l!(m+s)!(m+s-l)!}.
\label{fkfactors}
\end{eqnarray}

For a particular excited state $|m\rangle $ of the trap, one can prove that for certain values of
the detuning and of the Lamb-Dicke parameter, the sum in the last line of Eq.\ (\ref{fkfactors})
cancels out. If this occurs, and following Eq.\ (\ref{erates1d}), one can observe that this level
$|m\rangle $ is not emptied, and therefore becomes a dark state. The dark-state conditions for the
states $|1\rangle $ and $|2\rangle $ of the trap, are respectively:

\begin{mathletters}
\begin{eqnarray}
\label{dcond1}
\eta^{2}&=&s+1 \\
\label{dcond2}
\eta^{2}&=&(s+2)(1\pm (s+2)^{-1/2})
\end{eqnarray}
\end{mathletters}

Certainly, these conditions have some limitations. First of all, as the detuning must be integer
(i.e. $s$ must be an integer number), only particular values of $\eta$ are valid in order to
satisfy the dark-state conditions. This is not a major problem because the trap frequency can
be externally controlled. However, one  must note that for the case of one-atom problem these
conditions are certainly rather restrictive. Dynamical Cooling of a single atom is critically
sensitive to the deviations from the dark-state conditions. We show in \cite{Uns} that fortunately 
this is  not the case in the many atom problem with bosonic atoms. The  limitations are overcome
due to quantum statistical effects (bosonic enhancement effect).

\subsubsection{Scheme to cool into an excited state of the trap}
\label{sec:1DES}

Using the dark-state condition it is possible to design a cooling mechanism that allows the
confinement of the atom into an excited state of the trap. The cooling mechanism consists as in the
ground-state cooling in a sequence of several pulses:

\begin{itemize}
\item 1. Confinement pulse: exactly the same as in the ground-state cooling. The use of two
sligthly detuned confinement pulses is also recommended.

\item 2. Dark-State cooling pulse: which fullfills the dark-state condition for a selected excited
state.

\item 3. Auxiliary pulses that avoid the oscilatory character of the empty rates of the dark-state
cooling pulse, but do not affect the excited state in which the confinement is to be produced.
\end{itemize}

\begin{figure}[ht]
\begin{center}\
\epsfxsize=7.5cm
\hspace{0mm}
\psfig{file=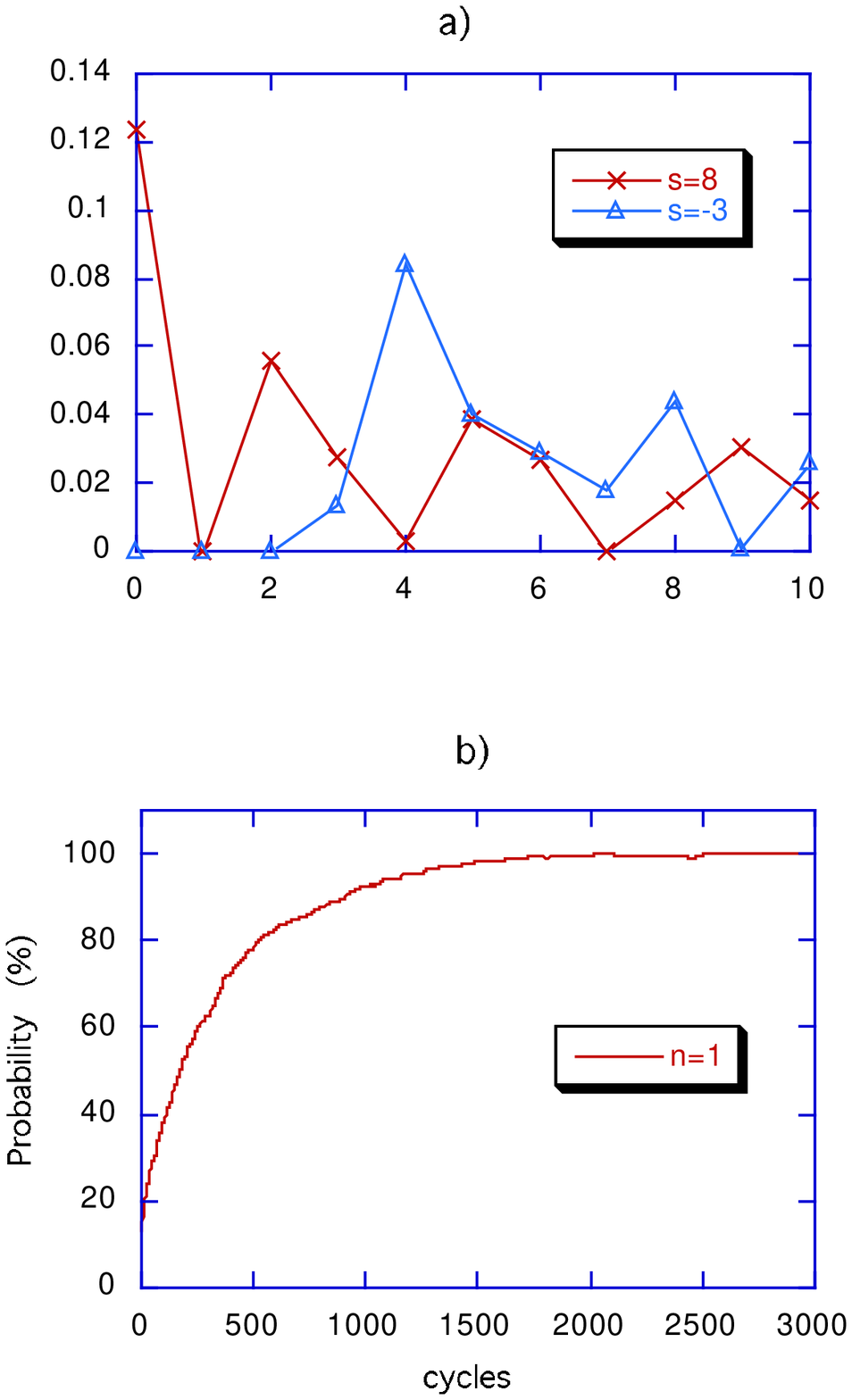,width=7.5cm}\\[0.3cm]
\caption{(a) Empty rates as a function of the trap level, for the case of $\eta =3.0$,
$\gamma = 0.01 \omega $, and a pulse with $\delta=s\omega$, with $s=8, -3$. (b)
Dynamics of the occupation probability of the level $n=1$ as a function of the applied cycles,
starting from the thermal state with the mean number $6$. Each cycle consists on a sequence of four
pulses of detunings
$\delta_{1,2,3,4}=-9\omega, 8\omega, -10\omega, -3\omega $ and durations $T_{1,2,3,4}=2\gamma/\Omega
^{2}$.}
\label{fig:3}
\end{center}
\end{figure}

In Figs.\ \ref{fig:3} the case of $\eta =3.0$, $\gamma =0.01\omega$ is analyzed. We have used cycles
of four pulses of detunings $\delta_{1,2,3,4}=-9\omega, 8\omega, -10\omega, -3\omega $ and durations
$T_{1,2,3,4}=2\gamma/\Omega ^{2}$ to confine the atom with almost total probability in the first
excited state of the trap. Pulses $1$ and $3$ are confinement pulses,  the pulse $2$ fulfills Eq.\
(\ref{dcond1}) and is therefore the dark-state pulse for the level $|1\rangle $, and the pulse
$4$ is an auxiliary pulse. In Fig.\ \ref{fig:3}a the empty rates corresponding to pulses $2$ and $4$
are depicted. Note that the both pulses do not empty  the  level $|1\rangle $, and that it is the
only level not emptied by them. Thus, the  level $|1\rangle $ becomes a trapping state. Fig.\
\ref{fig:3}b shows the dynamics of confinement in the state $|1\rangle $ for the considered
sequence of pulses, where the initial distribution was taken to be thermal with the mean $\langle n
\rangle =6$.

\begin{figure}[ht]
\begin{center}\
\epsfxsize=7.5cm
\hspace{0mm}
\psfig{file=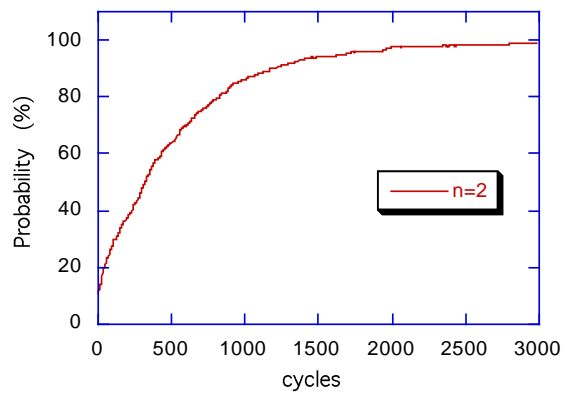,width=7.5cm}\\[0.3cm]
\caption{Dynamics of the occupation probability of the level $n=2$ as a function of the applied
cycles, starting from the thermal state with the mean number $6$. We consider the case of $\eta
=3.0$, $\gamma = 0.01 \omega $, and cycles of four pulses of detunings $\delta_{1,2,3,4}=-9\omega,
11\omega, -10\omega, -5\omega $ and durations $T_{1,2,3,4}=2\gamma/\Omega ^{2}$.}
\label{fig:4}
\end{center}
\end{figure}

The atom can also be confined in higher levels of the trap. Fig.\ \ref{fig:4} shows the dynamics of
the confinement into level $|2\rangle $, for the case of $\eta=3.065$, $\gamma =0.01\omega$, the
initial thermal distribution with the mean $\langle n \rangle =6$, and cycles of four pulses of
detunings
$\delta_{1,2,3,4}=-9\omega, 11\omega, -10\omega, -5\omega $ and durations
$T_{1,2,3,4}=2\gamma/\Omega ^{2}$. As previously, the pulses $1$ and $3$ are confinement pulses, the
pulse $2$ fulfills Eq.\ (\ref{dcond2}) and is therefore a dark-state pulse for the level $|2\rangle
$, and the pulse $4$ is an auxiliary pulse.

\section{Two Dimensions}
\label{sec:2D}

\subsection{Main new features in dimensions higher than one}
\label{sec:2Dnew}

In this section we consider the single atom cooling in two dimensions. The two dimensional problem
contains essentially all of the main new physical features that appear when one considers dimensions
higher than one, namely:

\begin{itemize}
\item The number of levels involved is larger  (squared for the two-dimensional case). Therefore the
atom can perform  a random walk in a larger phase space, which means that the cooling is neccesarly
slowlier than in the one-dimensional case.

\item Cooling in one direction means heating in the other directions. This fact makes neccesary the
use of lasers in different directions. Therefore, we will consider two lasers propagating in
directions $x$ and $y$, providing an electric field

\begin{equation}
\vec E=\vec E_{0}\left (e^{ikx}+Ae^{iky} \right )e^{-i\omega t} + c.c
\label{efield}
\end{equation}

where the parameter $A$ accounts for  a possible difference of intensities or dephasing between
both lasers. As we indicate below this parameter turns to be important under certain conditions.

\item Confinement pulses are no longer pulses of detuning $\delta = -\hat \eta ^{2} \omega  $. This
fact can be easily understood if we consider the case in which the atom absorbs a photon in
direction $x$. Assuming a detuning $\delta = -\hat \eta ^{2} \omega $, after an absorption the atom
makes a transition $|g,n_{x},n_{y}\rangle \rightarrow |e,n_{x}-{\cal O}(\eta ^{2}), n_{y}\rangle$,
and after the spontaneous emission  $|e,n_{x}-{\cal O}(\eta ^{2}), n_{y}\rangle \rightarrow
|g,n_{x}-{\cal O} (\eta ^{2}) \pm n_{x}', n_{y} \pm n_{y}'\rangle$, where both $n_{x}'$ and
$n_{y}'$ range from $0$ to ${\cal O} (\eta ^{2})$. Therefore the two-dimensional quantum number
$n=n_{x}+n_{y}$ changes into $n+n'$ where $n'$ ranges between ${\cal O} (-3\eta ^{2})$ and ${\cal
O} (\eta ^{2})$, and therefore an average heating is produced. For the two-dimensional problem,
the true confinement pulses are those with detuning $\delta = -2\hat \eta ^{2} \omega $ (in
general for $D$ dimensions $\delta = -D\hat \eta ^{2} \omega $), as can be easily proved by using
the same reasoning as above.

\item New cooling mechanisms can be designed, as we show below.

\end{itemize}

\subsection{Rate Equations}
\label{sec:2DRE}

The cooling dynamics is regulated by a set of rate equations as that of Eq.\ (\ref{rateeq1d}). The
transition rates in the two-dimensional laser arrangement considered are of the form:

\begin{eqnarray}
&& \Gamma_{\vec n \leftarrow \vec m}=\frac {\Omega^{2}}{2\gamma}\int_{0}^{2 \pi}d\phi\int_{0}^{\pi} \sin{\theta} d\theta {\cal W}(\phi , \theta )\nonumber \\ && \times
\Biggl | \sum_{l_{x},l_{y}=0}^{\infty}\frac {\gamma \langle n_{x}|e^{ik \sin{\theta} \cos{\phi} x}
|l_{x}\rangle \langle l_{x}|e^{ikx}|m_{x}\rangle}{[\delta-\omega (l_{x}-m_{x})]+i\gamma} \nonumber
\\
&& \qquad \times \langle n_{y}|e^{ik \sin{\theta}\sin{
\phi} x}|m_{y}\rangle \delta_{l_{y},m_{y}} \nonumber \\
&& + A\sum_{l_{x},l_{y}=0}^{\infty}\frac {\gamma \langle n_{y}|e^{ik \sin{\theta}
\sin{\phi} y}|l_{y}\rangle \langle l_{y}|e^{iky}|m_{y}\rangle}{[\delta-\omega
(l_{y}-m_{y})]+i\gamma} \nonumber \\
&& \qquad \times
\langle n_{x}|e^{ik \sin{\theta} \cos{\phi} x}|m_{x}\rangle \delta_{l_{x},m_{x}} \Biggr | ^{2}
\label{rates2d}
\end{eqnarray}

where $\vec n= (n_{x},n_{y})$. Proceeding as in the one-dimensional case, it is possible to obtain a
simple form for the empty rates:

\begin{eqnarray}
&& \Gamma_{\vec m} \simeq \frac{\Omega^{2}}{2\gamma} \left \{\left |\langle
m_{x}+s|e^{ikx}|m_{x}\rangle \right |^{2} \right \delimiter 0 \nonumber \\ && \left \delimiter 0
+|A|^{2}\left |\langle m_{y}+s|e^{iky}|m_{y}\rangle\right |^{2} \right \delimiter 0 \nonumber \\
&& \left \delimiter 0 +(A+A^{\ast})\langle m_{x}+s|e^{ikx}|m_{x}\rangle\langle
m_{y}+s|e^{iky}|m_{y}\rangle \delta_{s,0} \right \}
\label{erates2d}
\end{eqnarray}

\subsection{Destructive interference of the absorption amplitudes: New dark-states in two
dimensions}
\label{sec:2Dint}

Let us consider the case in which $\delta =0$, i.e. $s=0$. Let us also assume that both lasers in
directions $x$ and $y$ have the same intensity but are $\pi$-dephased, i.e. $A=-1$. Then, Eq.\
(\ref{erates2d}) becomes:

\begin{equation}
\Gamma_{\vec m} \simeq \frac{\Omega^{2}}{2\gamma} \left |\langle m_{x}+s|e^{ikx}|m_{x}\rangle
-\langle m_{y}+s|e^{iky}|m_{y}\rangle\right |^{2}
\label{erates2ds0}
\end{equation}

Eq.\ (\ref{erates2ds0}) shows that in the case of an isotropic trap, all the states in which
$m_{x}=m_{y}$, satisfy $\Gamma_{\vec m}=0$, i.e. are not emptied. Id est, the destructive
interference of the absorption of the two lasers produces that the diagonal levels become dark. As
we show in the following section, this fact can be employed to cool the atom into the ground state
of the two-dimensional trap with more efficiency than the simple extension of the one-dimensional
method to two dimensions. Moreover, if we set

\begin{equation}
A=-\frac {\langle m_{x}^{0}|e^{ikx}|m_{x}^{0}\rangle}{\langle m_{y}^{0}|e^{iky}|m_{y}^{0}\rangle}
\label{Am0n0}
\end{equation}

where $\vec m^{0}=(m_{x}^{0},m_{y}^{0})$ is a particular state of the two-dimensional trap, then
from Eq.\ (\ref{erates2d}) it is easy to observe that this particular level becomes a dark state.
Therefore, we can specifically design the factor $A$ to regulate the interference between the two
lasers in such a way that a selected state $\vec m^{0}$ becomes dark. As always, this fact can be
employed to design a cooling mechanism to pump the atom into the level $\vec m^{0}$. An example of
this selective cooling is discussed in section IV.E. Also one can use Eq.\ (\ref{Am0n0}) to
generalize the dark-state condition appearing from Eq.\ (\ref{erates2ds0}) for the more general
case of an anisotropic trap.

Additionally, the dark-state condition for the excited states presented in Sec.\ \ref{sec:1DEC},
remains valid for the two-dimensional case, allowing the cooling into excited states of the trap,
as we show in Sec.\ \ref{sec:2DE}.

\subsection{Cooling into the Ground State of the Trap}
\label{sec:2DG}

In this section we demonstrate how the previously presented destructive interference between the
lasers can be employed to improve the cooling into the ground state of the trap. As in
one dimension, the cooling mechanism consists of several pulses:

\begin{itemize}
\item 1. Confinement pulses: with the same function as in the one-dimensional case but, instead of
pulses with detuning $\delta = -\hat \eta^{2} \omega $, we use, following the reasonings of
Sec.\ \ref{sec:2Dnew}, $\delta = -2\hat \eta^{2} \omega $. The use of two sligthly detuned
confinement pulses is also needed.

\item 2. Dark-State cooling pulses: with detuning $\delta =0$, and with ratio between the laser
amplitudes $A=-1$. Following the discussion of the previous section the diagonal
states $m_{x}=m_{y}$ are dark for these pulses.

\item 3. Sideband cooling pulses: of course the dark-state pulses induce trapping in the diagonal
states, and not only in $(0,0)$. In order to avoid that, a usual Sideband cooling pulse is
neccesary, i.e. a pulse with detuning $\delta = -\omega $.

\item 4. Pseudo-confining pulses: as pointed out in Sec IV.A. the atom performs  its random walk in
a larger phase space, and therefore with the only use of pulses $1$, $2$ and $3$ the cooling is
neccesarly very slow. In order to improve the cooling time we use pulses with $\delta = -\hat \eta
^{2}$ and pulses with $\delta = -\hat \eta ^{2} /2$ (closest integer to this value). These pulses
pseudo-confine the atom below $n=\hat \eta ^{2}$ and $n=\hat \eta ^{2}/2$, respectively. Due to
this fact we call these pulses Pseudo-confinement pulses.
\end{itemize}

\begin{figure}[ht]
\begin{center}\
\epsfxsize=7.5cm
\hspace{0mm}
\psfig{file=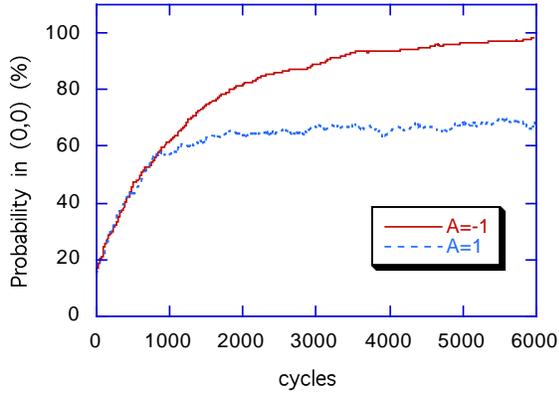,width=7.5cm}\\[0.3cm]
\caption{Dynamics of the occupation probability of the level $(0,0)$ as a function of the applied
cycles, starting from the thermal state with the mean number $6$. We consider the case of $\eta
=3.0$, $\gamma = 0.01 \omega $ and cycles of eight pulses of detunings
$\delta_{1,2,3,4,5,6,7,8}=-18\omega,-9\omega,-4\omega,0, -19\omega,-10\omega,-5\omega,-\omega $,
all of them with duration $T=2\gamma / \Omega^{2}$. In solid-line is represented the case of $A=-1$
for all the pulses, whereas in dashed-line the case $A=1$ is depicted.}
\label{fig:5}
\end{center}
\end{figure}

Fig.\ \ref{fig:5} shows the cooling dynamics for the case of $\eta =3.0$, $\gamma=0.01 \omega $, and
an initial thermal distribution with mean $\langle n \rangle =6$. We have used cycles of eight
pulses of detunings $\delta_{1,2,3,4,5,6,7,8}=-18\omega, -9\omega, -4\omega, 0, -19\omega,
-10\omega, -5\omega, -\omega $, all of them with duration $T=2\gamma / \Omega^{2}$. In solid-line is
represented the case of $A=-1$ for all the pulses, whereas in dashed-line the case $A=1$ is
depicted. Note that due to the form of Eq.\ (\ref{erates2d}), only for $\delta =0$ it is relevant
if $A=1$ or $A=-1$. Pulses $1$ and $5$ are confinement pulses. Pulse $4$ is the dark-state pulse
(for the case $A=-1$). Pulse $8$ is the Sideband Cooling pulse. Pulses $2$, $3$, $6$ and $7$ are
pseudo-confinement pulses. One can observe that for the case of $A=-1$ the cooling is very much
effective than for the case of $A=1$. Therefore, the interference mechanism which appears in two
dimensions (and in general in dimensions higher than one) between the absorption amplitudes  of
lasers in different directions can be employed to improve the cooling technique.

\subsection{Cooling into an Excited State of the Trap}
\label{sec:2DE}

In this section, we employ the two different kinds of dark-states, i.e. those produced by the
special form of the Franck-Condon factors (Sec.\ \ref{sec:1DEC}), and those produced by the
interference of lasers in different directions (Sec.\ \ref{sec:2Dint}), to confine the atom into an
excited state of the harmonic trap. As in the previous section the use of confinement and
pseudo-confinement pulses is in this case also neccesary.

\begin{figure}[ht]
\begin{center}\
\epsfxsize=7.5cm
\hspace{0mm}
\psfig{file=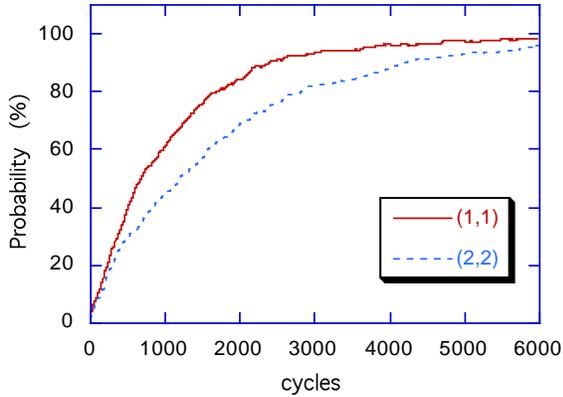,width=7.5cm}\\[0.3cm]
\caption{In solid line, the dynamics of the occupation probability of the level $(1,1)$ is depicted
as a function of the applied cycles, starting from the  thermal state with the mean number $6$. We
have considered the case of $\eta =3.0$, $\gamma=0.01\omega$, and cycles of eight pulses of
detunings
$\delta_{1,2,3,4,5,6,7,8}=-18\omega,-9\omega,-4\omega,8\omega,-19\omega,-10\omega,-5\omega,-3\omega
$, all of them with duration $T=2\gamma / \Omega^{2}$. In dashed line, the dynamics of the
occupation probability of the level $(2,2)$ is shown for the case of cycles of seven pulses of
detunings $\delta_{1,2,3,4,5,6,7}=-18\omega,-9\omega,-4\omega,11,-19\omega,-10\omega,-5\omega$, all
of them with duration $T=2\gamma / \Omega^{2}$.}
\label{fig:6}
\end{center}
\end{figure}

Let us analyse in first place the cooling using the Franck-Condon properties. Fig.\ \ref{fig:6}
shows in solid line the dynamics of the probability to occupy the level $(1,1)$ of the trap, for
the case of $\eta =3.0$, $\gamma =0.01 \omega $, the initial thermal distribution with the mean
$\langle n \rangle =6$, and a sequence of pulses of detunings: $\delta_{1,2,3,4,5,6,7,8}=
-18\omega,-9\omega,-4\omega,8\omega,-19\omega,-10\omega,-5\omega,-3\omega $, all of them with
duration $T=2\gamma / \Omega^{2}$. As previously, pulses $1$ and $5$ are confinement
pulses, and  pulses $2$, $3$, $6$ and $7$ are pseudo-confinement pulses. Pulse $4$ fullfills
Eq.\ (\ref{dcond1}), and from Eq.\ (\ref{erates2d}) it is trivial to see that for an isotropic trap
the level $(1,1)$ of the trap becomes a dark-state for this pulse. Finally pulse $8$ avoid
the formation of undesired unemptied states. One can observe that an almost complete confinement of
the atom into the level $(1,1)$ can be obtained.

Fig.\ \ref{fig:6} shows in dashed line the dynamics of the probability to occupy the level $(2,2)$
of the trap, for the case of $\eta =3.065$, $\gamma =0.01 \omega $ and a sequence of pulses of
detunings:
$\delta_{1,2,3,4,5,6,7}= -18\omega,-9\omega,-4\omega,11\omega,-19\omega,-10\omega,-5\omega $, all of
them with duration $T=2\gamma / \Omega^{2}$. The pulses play the same role as in the previous
case. One can observe that the atom can be confined with large probability into
the level $(2,2)$.

\begin{figure}[ht]
\begin{center}\
\epsfxsize=7.5cm
\hspace{0mm}
\psfig{file=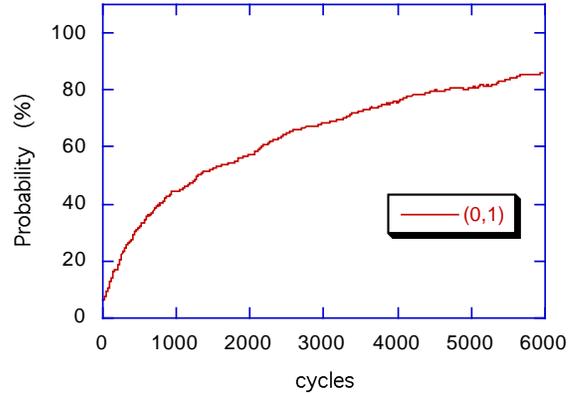,width=7.5cm}\\[0.3cm]
\caption{Dynamics of the occupation probability of the level $(0,1)$ as a function of the applied
cycles, starting from a thermal state with a mean number $6$. We consider the case of $\eta =3.0$,
$\gamma = 0.01 \omega $, and cycles of eight pulses of detunings
$\delta_{1,2,3,4,5,6,7,8}=-18\omega,-9\omega,-4\omega,0, -19\omega,-10\omega,-5\omega,-\omega $,
all of them with duration $T=2\gamma / \Omega^{2}$, except pulse $4$ which has a duration
$T=8\gamma / \Omega^{2}$ All of them have $A=-1$ except pulse $4$ which has $A=1/8$.}
\label{fig:7}
\end{center}
\end{figure}

Finally, let us consider the cooling mechanism using the dark-states produced by the interference
between the two lasers. Fig.\ \ref{fig:7} shows the cooling dynamics for the case of $\eta =3.0$,
$\gamma =0.01 \omega $, the  initial thermal distribution with the mean $\langle n \rangle =6$, and
a sequence of pulses of detunings:
$\delta_{1,2,3,4,5,6,7,8}=-18\omega,-9\omega,-4\omega,0,-19\omega,-10\omega,-5\omega, -2\omega
$, all of them with duration $2\gamma / \Omega^{2}$, except pulse $4$ which has a duration $8\gamma
/\Omega^{2}$. For all the pulses we consider the case of $A=-1$, except pulse $4$ for which we use
$A=1/8$, satisfying Eq.\ (\ref{Am0n0}) for the level $(0,1)$. As previously, pulses $1$ and $5$
are confinement pulses, and  pulses $2$, $3$, $6$ and $7$ are pseudo-confinement pulses. Pulse $4$
is a dark-state pulse for the level $(0,1)$, and pulse $8$ avoids the formation of undesired
unemptied states. One can observe that the atom can be confined with large probability into the
level $(0,1)$.

\subsection{Limitations of the dark-state produced by the interference of absorption amplitudes}
\label{sec:2Dlim}

We have shown that the destructive interference between the lasers in both directions allows
the improvement of the cooling in the ground state of the harmonic trap, and also provides a new
mechanism to cool the atoms in a selected excited state of the trap. Unfortunately this
interference mechanism is not valid for large values of the Lamb-Dicke parameter $\eta$. The reason
is the following. The interferential mechanism is based on pulses with detuning $\delta =0$, i.e.
pulses whose resonant absorption produces a jump between one state of the trap and itself. For
sufficiently large $\eta$ the Debye-Waller factor $e^{-\eta^{2}/2}$, which appears in the Franck-Condon factors (Eq.\ (\ref{fkfactors})), becomes very small, and therefore the Franck-Condon
factors connecting one state with itself are very small. This fact implies that the empty rates
practically vanish for all the states, i.e. all the states of the trap become dark, whatever the
value of $A$, and therefore the interference mechanism is no more valid. The interference mechanism
can be employed, roughly speaking, for  $\eta$ not larger than 4.

\section{Conclusions}
\label{sec:Conclusions}

The cooling dynamics of a single atom trapped in one-dimensional and two-dimensional harmonic
potentials has been analysed. Using a particular sequence of confinement plus fine-cooling pulses
the atom can be cooled into the ground state of the one-dimensional trap. Moreover, by using the
special properties of the Franck-Condon factors, a selected excited state of the trap can be made
dark, and then a cooling mechanism can be designed to cool into it. We have determined the
dark-state condition, and proved via Monte Carlo simulations that the atom can be confined with
almost total probability in the selected excited state. However, the formation of the dark-state is
very sensititive to the deviations from the dark-state condition. This condition is by far very much
less restrictive for the case of many atoms in the trap as we show in \cite{Uns}.

The difficulties to cool in higher dimensions han been analysed on studying the two-dimensional
case. In particular one needs a new laser arrangement with two orthogonal lasers and confinement
pulses with twice the detuning of the one-dimensional confinement pulses. The difference between the
amplitudes of the orthogonal lasers acts as a new degree of freedom that can be used for Lamb-Dicke
parameters $\eta < 4$ to produce a destructive interference between the absorption amplitudes of
both lasers. This destructive interference can be used to create dark-states which, as we have
shown numerically, allow the improvement of the efectivity of the cooling into the ground-state.
Also this interference mechanism can be used to cool into selected excited states of the trap.

For larger Lamb-Dicke parameters $(\eta > 4)$ the interference method is no longer valid. In that
regime, the only possibility seems to be the extension of the one-dimensional results of
\cite{Morigi} (basically based on the use of blue-detuned lasers to empty the states different than
the ground-state) to higher dimensions, by using a two-laser arrangement, and confinement and
pseudo-confinement pulses. Fortunately, the dark-state condition based on the properties of the
Franck-Condon factors remains effective for large Lamb-Dicke parameters, allowing the excited-state
cooling also in this regime.

\bigskip

{\bf Acknowledgements} Partial support from the Spanish Direcci\'on General de Investigaci\'on
Cient\'\i fica y t\'ecnica (Grant No. PB95-0955) and from the Junta de Castilla y Le\'on
(Grant No SA 81/96) is acknowledged. L.S. wants to thank M. L. for hospitality during his stay in
the Institut f\"ur Theoretische Physik of Hannover. This work has been supported by the Deutsche
Forschungsgemeinschaft under SFB  407.

\end{document}